\documentclass[twocolumn,aps,a4paper,numerical,prl,reprint, superscriptaddress,floatfix, nofootinbib]{revtex4-1} 

\usepackage{graphicx}
\usepackage{dcolumn}
\usepackage{bm}
\begin{document}



\title{It's not the voting that's democracy, it's the counting: statistical detection of systematic election irregularities}

\author{Peter Klimek}
\affiliation{Section for Science of Complex Systems; Medical University of Vienna; Spitalgasse 23; A-1090; Austria}
\author{Yuri Yegorov}
\affiliation{Institut f\"ur Betriebswirtschaftslehre; University of Vienna; Br\"unner Stra\ss e 72; 1210 Vienna; Austria}
\author{Rudolf Hanel}
\affiliation{Section for Science of Complex Systems; Medical University of Vienna; Spitalgasse 23; A-1090; Austria}
\author{Stefan Thurner}
\affiliation{Section for Science of Complex Systems; Medical University of Vienna; Spitalgasse 23; A-1090; Austria}
\affiliation{Santa Fe Institute; 1399 Hyde Park Road; Santa Fe; NM 87501; USA}
\affiliation{IIASA, Schlossplatz 1, A-2361 Laxenburg; Austria}

\begin{abstract} Democratic societies are built around the principle of free and fair elections, that each citizen's vote should count equal.
National elections can be regarded as large-scale social experiments, where people are grouped into usually large numbers 
of electoral districts and vote according to their preferences.
The large number of samples implies certain statistical consequences for the polling results 
which can be used to identify election irregularities. 
Using a suitable data collapse, we find that vote distributions of elections with alleged fraud show a kurtosis of hundred times more than normal elections on certain levels of data aggregation.
As an example 
we show that reported irregularities in recent Russian elections are indeed well explained by systematic ballot stuffing
and develop a parametric model quantifying to which extent fraudulent mechanisms are present.
We show that if specific statistical properties are present in an election, the results do not represent the will of the people.
We formulate a parametric test detecting these statistical properties in election results.
Remarkably, this technique produces similar outcomes irrespective of the data resolution and thus allows for cross-country comparisons. \end{abstract}

\maketitle
Free and fair elections are the cornerstone of every democratic society \cite{Diamond06}.
A central characteristic of elections being free and fair is that each citizen's vote counts equal.
However, already Joseph Stalin believed that 
"It's not the people who vote that count; It's the people who count the votes."
How can it be distinguished whether an election outcome represents the will of the people or the will of the counters?

Elections can be seen as large-scale social experiments.
A country is segmented into a usually large number of electoral units. 
Each unit represents a standardized experiment where each citizen articulates his/her political preference via a ballot.
Although elections are one of the central pillars of a fully functioning democratic process,
relatively little is known about how election fraud impacts and corrupts the results of these standardized experiments \cite{Lehoucq03, Alvarez08}.

There is a plethora of ways of tampering with election outcomes, 
for instance the redrawing of district boundaries known as gerrymandering, or the barring of certain demographics from their right to vote.
Some practices of manipulating voting results leave traces which may be detected by statistical methods.
Recently, Benford's law \cite{Benford38} experienced a renaissance as a potential election fraud detection tool \cite{Mebane06}.
In its original and naive formulation, Benford's "law" is the observation that for many real world processes the logarithm of the first significant digit is uniformly distributed.
Deviations from this law may indicate that other, possibly fraudulent mechanisms are at work.
For instance, suppose a significant number of reported vote counts in districts is completely made up and invented by someone preferring to pick numbers which are multiples of ten.
The digit "0" would then occur much more often as the last digit in the vote counts when compared to uncorrupted numbers.
Voting results from Russia \cite{Mebane09}, Germany \cite{Breunig11}, Argentina \cite{Cantu11} and Nigeria \cite{Beber12} have been tested for the presence of election fraud using variations of this idea of digit-based analysis.
However, the validity of Benford's law as a fraud detection method is subject to controversy \cite{Deckert11, Mebane11}.
The problem is that one needs to firmly establish a baseline of what the {\it expected} distribution of digit occurrences for fair elections should be.
Only then it can be asserted if {\it actual} numbers are over- or underrepresented and thus suspicious.
What is missing in this context is a theory that links specific fraud mechanisms to statistical anomalies \cite{Deckert11}.

A different strategy for detecting signals of election fraud is to look at the distribution of vote and turnout numbers as in \cite{Mebane04}.
This has been extensively done for the Russian presidential and Duma elections over the last 20 years \cite{Sukhovolsky94, Myagkov08, Myagkov09}.
These works focus on the task of detecting two mechanisms, the stuffing of ballot boxes and the reporting of contrived numbers.
It has been noted that these mechanisms are able to produce different features of vote and turnout distributions than those observed in fair elections.
While for Russian elections between 1996 and 2003 these features were "only" observed in a relatively small number of electoral units,
they eventually spread and percolated through the entire Russian federation from 2003 onwards.
According to Myagkov and Ordeshook \cite{Myagkov08} "[o]nly Kremlin apologists and Putin sycophants argue that Russian elections meet the standards of good democratic practice".
This point was further substantiated with election results from the 2011 Duma and 2012 presidential elections \cite{Shpilkin09, Shpilkin11, Shpilkin12}.
Here it was also observed that ballot stuffing not only changes the shape of vote and turnout distributions, but also induces a high correlation between them.
Unusually high vote counts tend to {\it co-occur} with unusually high turnout numbers.

\begin{figure*}[btp]
 \begin{center}
 \includegraphics[width=140mm]{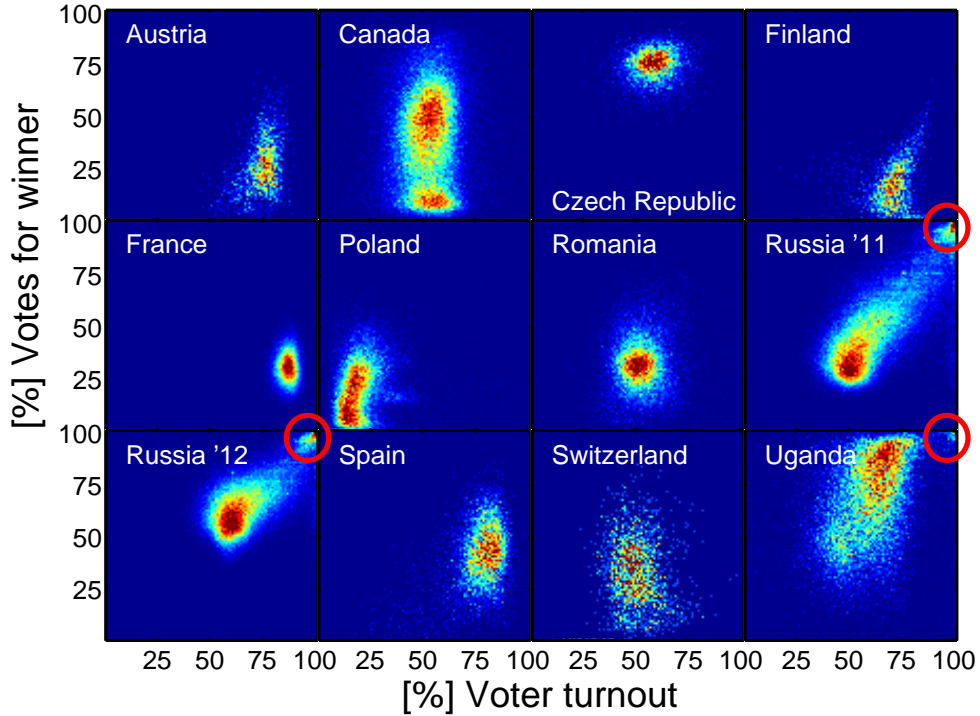}
   \end{center}
 \caption{Election fingerprints: 2-d histograms of the number of units for a given 
voter turnout (x-axis) and the percentage of votes (y-axis) for the winning party (or candidate) in recent elections from different countries 
(from left to right, top to bottom: Austria, Canada, Czech Republic, Finland, France, Poland, Romania, Russia '11 and '12, Spain, Switzerland, Uganda) are shown.
Color represents the number of units with corresponding vote and turnout numbers. 
The units usually cluster around a given turnout and vote percentage level.
In Uganda and Russia these clusters are 'smeared out' to the upper right region of the plots, reaching a second peak at a 100\% turnout and a 100\% of votes (red circles).
In Canada there are clusters around two different vote values, corresponding to the {\it Qu\'eb\'ecois} and English Canada (see SI).
In Finland the main cluster is smeared out into two directions (indicative of voter mobilization due to controversies surrounding the True Finns).}
  \label{Figure1}
 \end{figure*}

Several recent advances in the understanding of statistical regularities of voting results are due to the 
application of statistical physics concepts to quantitative social dynamics \cite{Castellano09}.
In particular several approximate statistical laws of how vote and turnout are distributed have been identified \cite{Costa03, Lyra03, Mantovani11},
some of them are shown to be valid across several countries \cite{Fortunato07, Borghesi10}.
It is tempting to think of deviations from these approximate statistical laws as potential indicators for election irregularities which are valid cross-nationally.
However, the magnitude of these deviations may vary from country to country due to different numbers and sizes of electoral districts.
Any statistical technique quantifying election anomalies across countries should not depend on the size of the underlying sample nor its aggregation level, i.e. the size of the electoral units.
As a consequence, a conclusive and robust signal for a fraudulent mechanism, e.g. ballot stuffing, must not disappear if the same dataset is studied on different aggregation levels.

In this work we expand earlier work on statistical detection of election anomalies in two directions.
First, we test for reported statistical features of voting results (and deviations thereof) in a cross-national setting, and discuss their dependence on the level of data aggregation.
As the central point of this work we propose a parametric model to statistically quantify to which extent fraudulent processes, such as ballot stuffing, may have influenced the observed election results.
Remarkably, under the assumption of coherent geographic voting patterns \cite{Borghesi10, Agnew96}, the parametric model results do not depend significantly on the aggregation level of the election data or the size of the data sample.

\section{Data and Methods}

\subsection{Election data}

Countries were selected by data availability.
For each country we require availability of at least one aggregation level where the average population per territorial unit $\bar n_{pop} \leq 5000$.
This limit for $\bar n_{pop}$ was chosen to include a large number of countries, that have a comparable level of data resolution.
We use data from recent parliamentary elections in Austria, Canada, Czech Republic, Finland, Russia (2011), Spain and Switzerland, the European Parliament elections in Poland 
and presidential elections in the France, Romania, Russia (2012) and Uganda.
Here we refer by "unit" to any incarnation of an administrative boundary (such as districts, precincts, wards, municipals, provinces, etc.) of a country on any aggregation level.
If the voting results are available on different levels of aggregation, we refer to them by Roman numbers, i.e. Poland-I refers to the finest aggregation level for Poland, Poland-II to the second finest, and so on.
For each unit on each aggregation level for each country we have the data of the number of eligible persons to vote, valid votes and votes for the winning party/candidate.
Voting results were obtained from official election homepages of the respective countries, for more details see SI Tab.S1.
Units with an electorate smaller than 100 are excluded from the analysis, to prevent extreme turnout and vote rates as artifacts from very small communities.
We tested robustness of our findings with respect to the choice of a minimal electorate size and found that the results do not significantly change if the minimal size is set to 500.

The histograms for the 2d-vote-turnout distributions (vtds) for the winning parties, also referred to as "fingerprints", are shown in Fig.\ref{Figure1}.

\begin{figure}[tbp]
 \begin{center}
 \includegraphics[width=87mm]{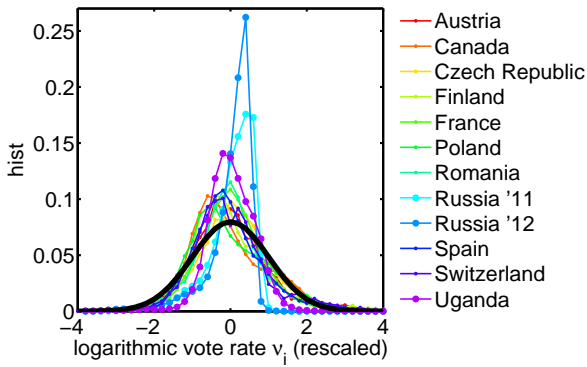}
  \end{center}
 \caption{A simple way to compare data from different elections in different countries on a similar aggregation level is to present the distributions of 
the logarithmic vote rates $\nu_i$ of the winning parties as rescaled distributions with zero-mean and unit-variance \cite{Borghesi10}.
Large deviations from other countries can be seen for Uganda and Russia with the plain eye. For more detailed results see Tab.S3.}
  \label{SIFigureColl}
 \end{figure}


\subsection{Data collapse}

It has been shown that by using an appropriate re-scaling of election data, the distributions of votes and turnouts approximately follow a Gaussian \cite{Borghesi10}.
Let $W_i$ be the number of votes for the winning party and $N_i$ the number of voters in any unit $i$.
A re-scaling function is given by the {\it logarithmic vote rate}, $\nu_i = \log \frac{N_i-W_i}{W_i}$ \cite{Borghesi10}.
In units where $W_i\geq N_i$ (due to errors in counting or fraud) or $W_i=0$ $\nu_i$ is not defined, and the unit is omitted in our analysis.
This is a conservative definition, since districts with extreme but feasible vote and turnout rates are neglected (for instance, in Russia 2012 there are 324 units with 100\% vote and 100\% turnout).

\begin{figure*}[tbp]
 \begin{center}
 \includegraphics[width=120mm]{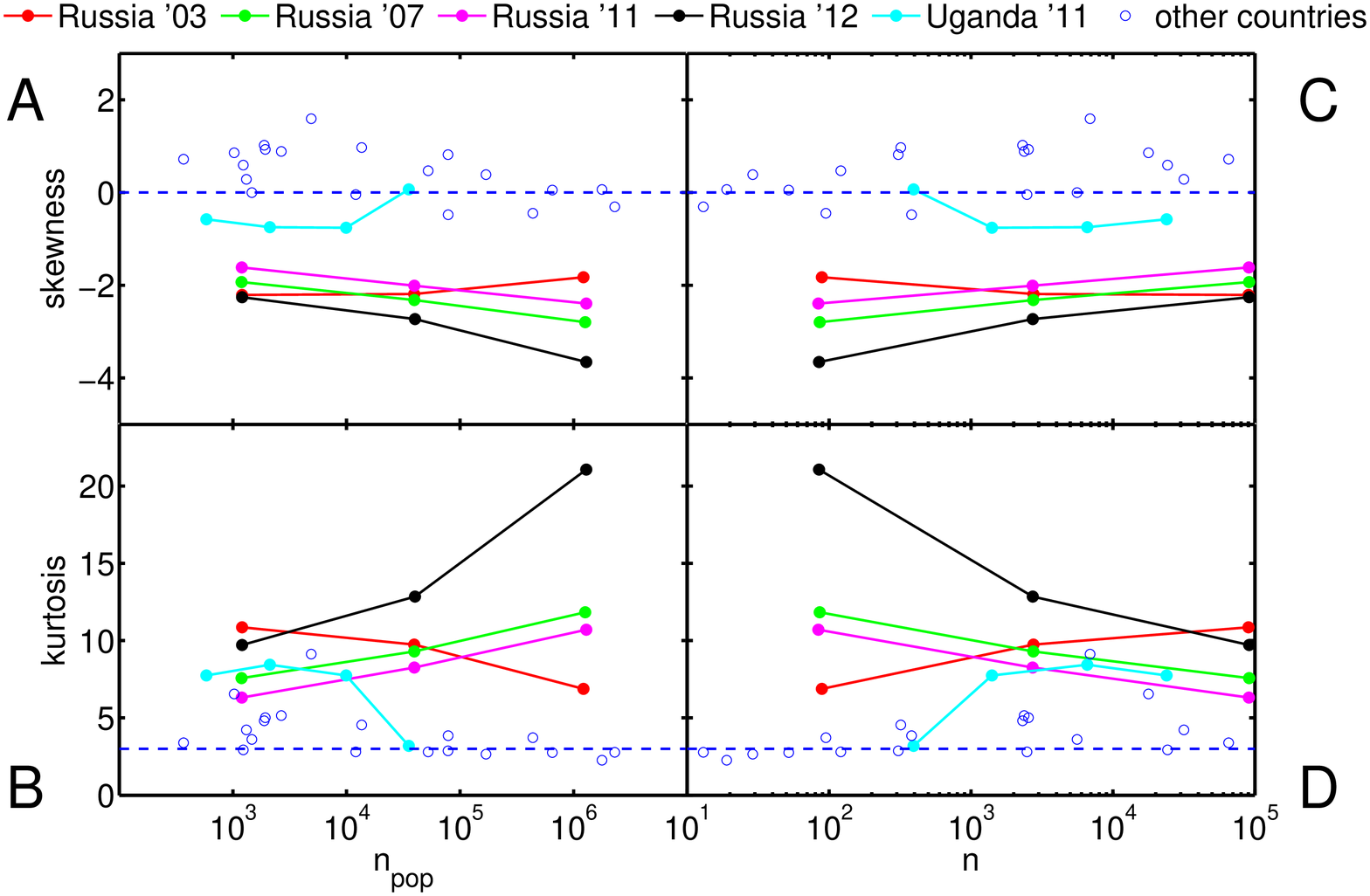}
  \end{center}
 \caption{ For each country on each aggregation level skewness and kurtosis of the logarithmic vote rate distributions are shown as a function of the average electorate per unit $\bar n_{pop}$ and the number of units $n$. Results for Russia and Uganda are highlighted. The values for all other countries cluster around 0 and 3, which are the values expected for normal distributions. On the largest aggregation level election data from Uganda and Russia can not be distinguished from other countries.}
  \label{MomentAggregate}
 \end{figure*}

\subsection{Parametric model}

To motivate our parametric model for the vtd,
observe that the vtd for Russia and Uganda in Fig.\ref{Figure1} are clearly bimodal, both in turnout and votes. 
One cluster is at intermediate levels of turnout and votes.
Note that it is smeared towards the upper right parts of the plot.
The second peak is situated in the vicinity of the 100\% turnout, 100\%  votes point. 
This suggests two modes of fraud  mechanisms being present, {\em incremental} and {\em extreme} fraud.
Incremental fraud means that with a given 
rate ballots for one party are added to the urn and votes for other parties are taken away.
This occurs within a fraction $f_i$ of units.
In the election fingerprints in Fig.\ref{Figure1} these units are those associated with the smearing to the upper right side.
Extreme fraud corresponds to reporting a complete turnout and almost all votes for a single party. 
This happens in a fraction $f_e$ of units.
These form the second cluster near 100\% turnout and votes for the winning party.

\begin{figure}[tbp]
 \begin{center}
 \includegraphics[width=87mm]{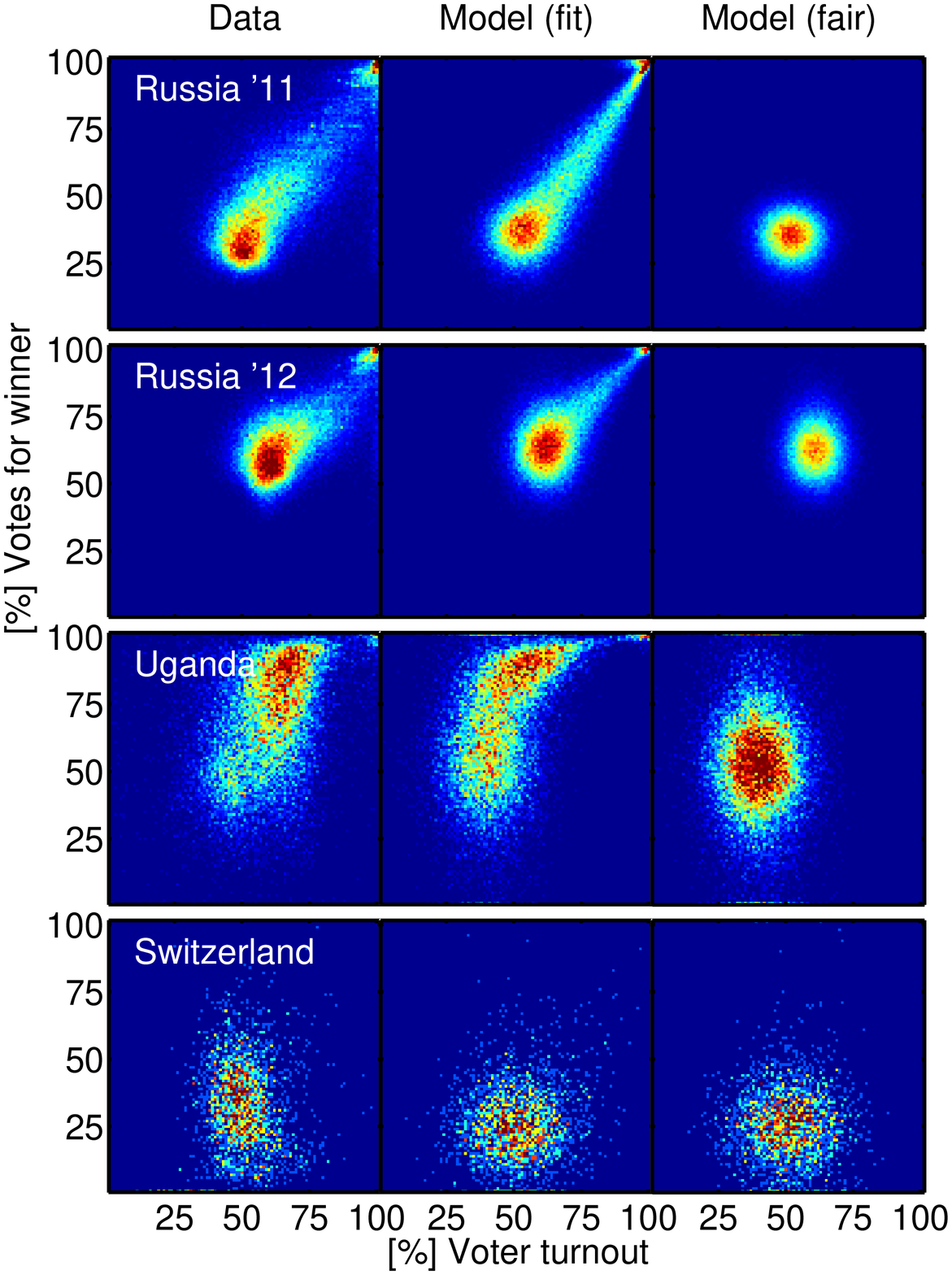}
  \end{center}
 \caption{Comparison of observed and modeled vtds for (top to bottom) Russia 2011, Russia 2012, Uganda and Switzerland.
The left column shows the observed election fingerprints, the middle column shows a fit with the fraud model.
The column to the right shows the expected model outcome of fair elections (i.e. absence of fraudulent mechanisms $f_i=f_e=0$). 
For Switzerland the fair and fitted model are almost the same.
The results for Russia and Uganda can be explained by the model assuming a large number of fraudulent units.}
  \label{Figure2}
 \end{figure}

For simplicity we assume that within each unit turnout and voter preferences  
can be represented by a Gaussian distribution with the mean and standard deviation taken from the actual sample, see SI Fig.S1.
This assumption of normality is not valid in general.
For example the Canadian election fingerprint of Fig.\ref{Figure1} is clearly bimodal in vote preferences (but not in turnout).
In this case, the deviations from approximate Gaussianity are due to a significant heterogeneity within the country.
In the particular case of Canada this is known to be due to the mix of the anglo- and francophone population.
Normality of the observed vote and turnout distributions is discussed in the SI, see Tab.S2.

Let $V_i$ be the number of valid votes in unit $i$.
The first step in the model is to compute the empirical turnout distribution, $V_i/N_i$, and the empirical vote distribution, $W_i/N_i$, over all units from the election data.
To compute the {\it model} vtd the following protocol is then applied to each unit $i$.
\begin{itemize}
\item For each $i$, take the electorate size $N_i$ from the election data.
\item Model turnout and vote rates for $i$ are drawn from normal distributions. The mean of the model turnout (vote) distribution is estimated from the election data as the value that maximizes the empirical  turnout (vote) distribution.  The model variances are also estimated from the width of the empirical distributions, see SI and Fig.S1 for details.
\item {\it Incremental fraud}. With probability $f_i$ ballots are taken away from both the non-voters and the opposition and are added to the winning party's ballots. The fraction of ballots which are shifted to the winning party is again estimated from the actual election data.
\item {\it Extreme fraud}. With probability $f_e$ almost all ballots from the non-voters and the opposition are added to the winning party's ballots.
\end{itemize}
The first step of the above protocol ensures that the actual electorate size numbers is represented in the model. 
The second step guarantees that the overall dispersion of vote and turnout preferences of the country's population are correctly represented in the model.
Given nonzero values for $f_i$ and $f_e$, incremental and extreme fraud are then applied in the third and fourth step, respectively.
For a complete specification of these fraud mechanisms, see the SI.

\subsection{Estimating the fraud parameters}

Values for $f_i$ and $f_e$ are reverse engineered from the election data in the following way.
First, model vtds are generated according to the above scheme, for each combination of $(f_i,f_e)$ values, where $f_i$ and $f_e \in \{0, 0.01, 0.02, \dots 1\}$.
We then compute the point-wise sum of the square difference of model and observed vote distributions for each pair $(f_i,f_e)$ and extract
the pair giving the minimal difference.
This procedure is repeated for 100 iterations, leading to 100 pairs of fraud parameters $(f_i, f_e)$.
In the following we report the average values of these $f_i$ and $f_e$ values, respectively, and their standard deviations.
For more details see SI. 

\begin{figure*}[tbp]
 \begin{center}
 \includegraphics[width=120mm]{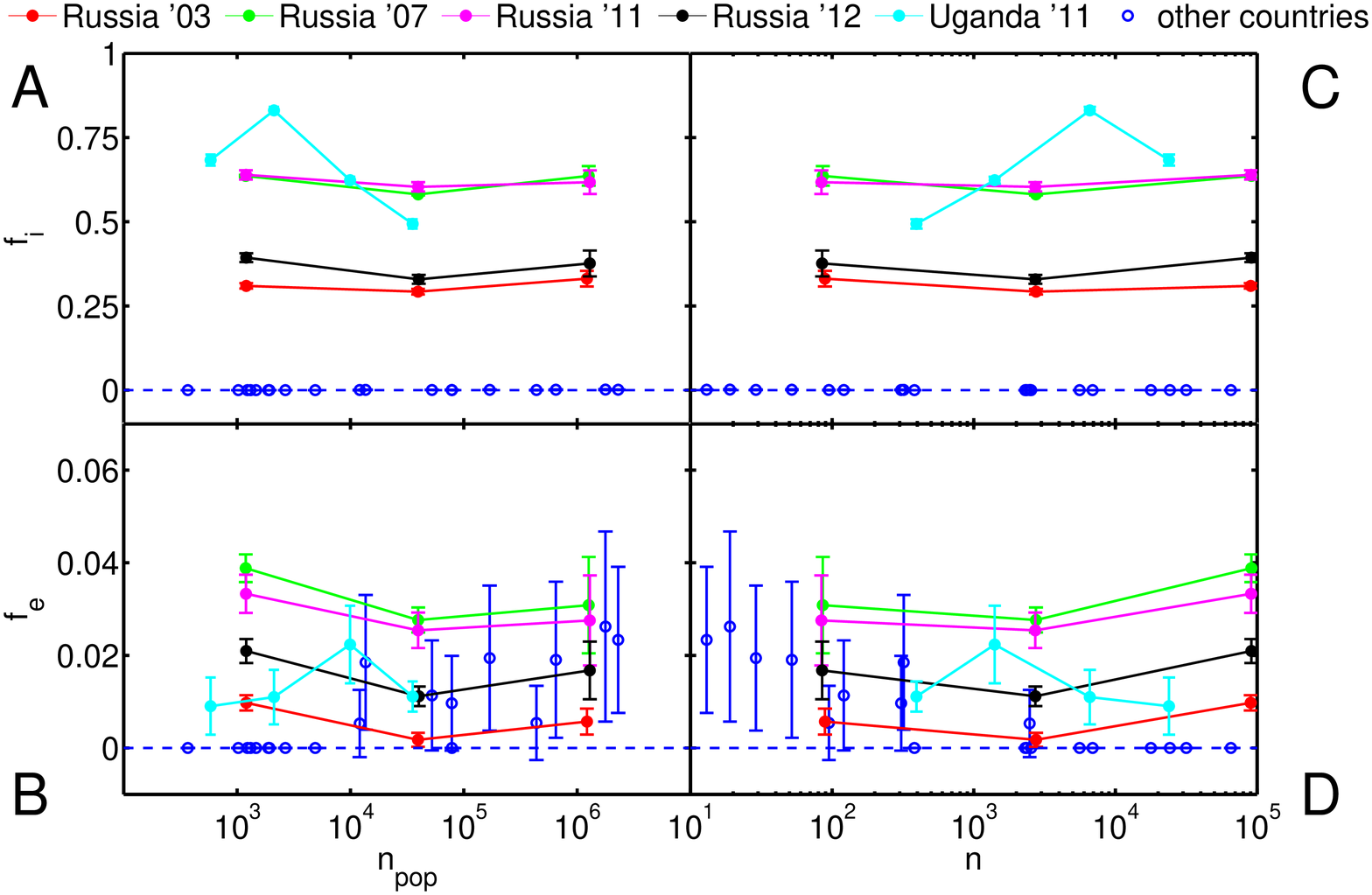}
  \end{center}
 \caption{ For each country on each aggregation level values for $f_i$ and $f_e$ as given in Tab.S3 are shown as a function of the average electorate per unit $\bar n_{pop}$ and the number of units $n$. Results for Russia and Uganda are  highlighted. The values for all other countries are close to zero, indicating that the data is best described by the absence of the ballot stuffing mechanism. Parameter values for $f_i$ and $f_e$ for Russia and Uganda remain significantly above zero for all aggregation levels. Note that in D the error margins for $f_e$ values in the range $10<n<100$ (as well as for the corresponding values $f_e$ in C) get increasingly large, whereas $f_i$ estimates in this range stay robust.}
  \label{FraudAggregate}
 \end{figure*}

\begin{figure}[tbp]
 \begin{center}
 \includegraphics[width=87mm]{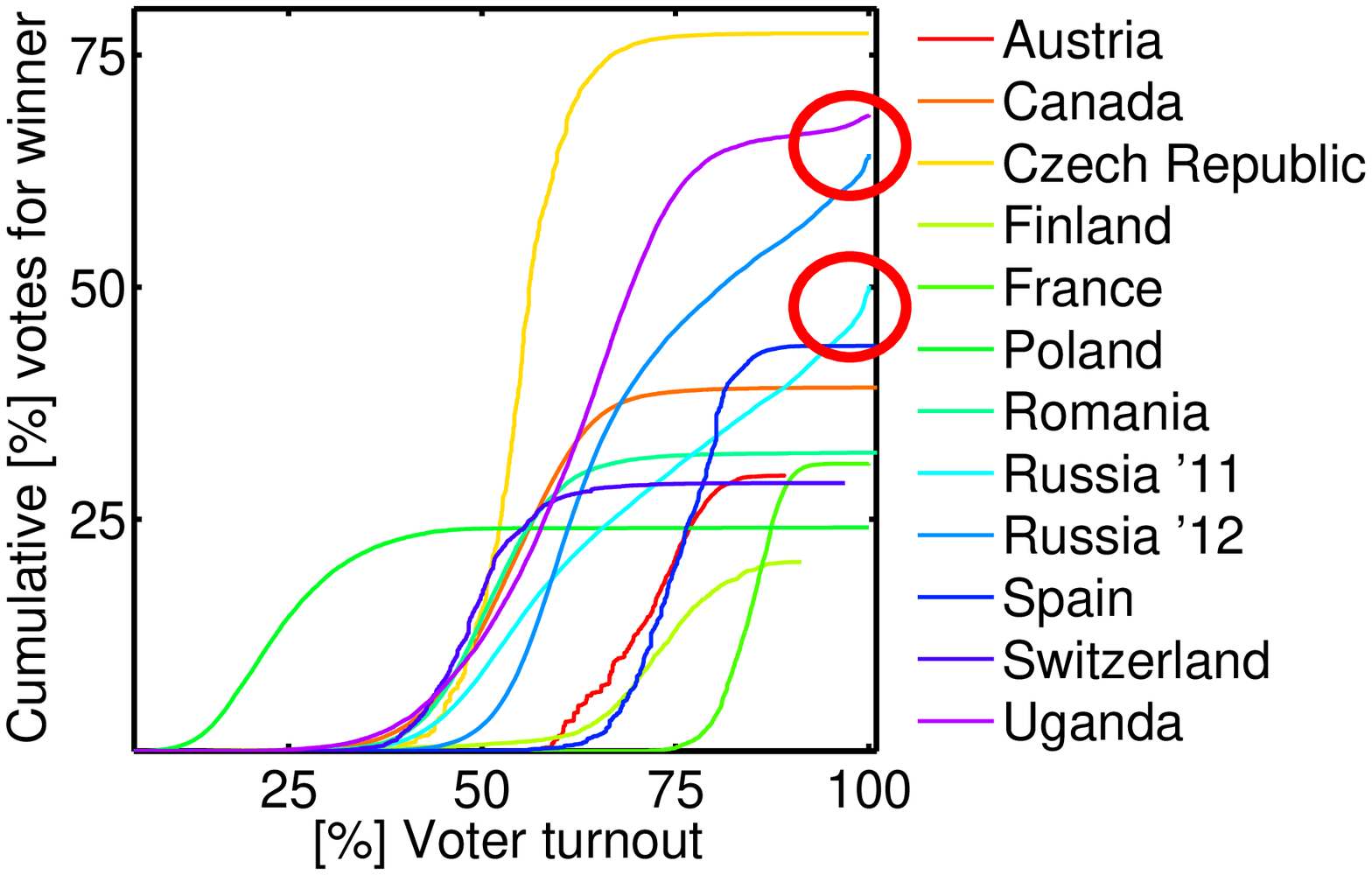}
  \end{center}
 \caption{The ballot stuffing mechanism can  be visualized by considering the cumulative number of votes as a function of turnout.
Each country's election winner is represented by a curve which  typically takes the shape of a sigmoid function reaching a plateau.
In contrast to the other countries, Russia and Uganda do not tend to develop this plateau but instead show a pronounced increase 
(boost) close to complete turnout. Both irregularities are indicative of the two ballot stuffing modes being present.}
  \label{Figure3}
 \end{figure}

\section{Results}

\subsection{Fingerprints}

Fig.\ref{Figure1} shows 2-d histograms (vtds) for the number of units for a given fraction of voter turnout  (x-axis)
and for the percentage of votes  for the winning party (y-axis).
Results are shown for Austria, Canada, Czech Republic, Finland, France, Poland, Romania, Russia, Spain, Switzerland and Uganda.
For each of these countries the data is shown on the finest aggregation level, where $\bar n_{pop} \leq 5000$.
These figures can be interpreted as fingerprints of several processes and mechanisms leading to the overall election results.
For Russia and Uganda the shape of these fingerprints differ strongly from the other countries. 
In particular there is a  large number of territorial units (thousands) with approximately 100\% percent turnout and at the same time 
about 100 \% of votes for the winning party.


\subsection{Approximate normality}

In Fig.\ref{SIFigureColl} we show the distribution of $\nu_i$ for each country. 
Roughly, to first order the data from different countries collapse to 
an approximate Gaussian, as previously observed \cite{Borghesi10}. 
Clearly, the data for Russia falls out out of line. 
Skewness and kurtosis for the distributions of $\nu_i$ are listed for each data-set and aggregation level in Tab.S3.
Most strikingly, the kurtosis of the distributions for Russia (2003, 2007, 2011 and 2012) exceed the kurtosis of each other country on the coarsest aggregation level by a factor of two to three.
Values for the skewness of the logarithmic vote rate distributions for Russia are also persistently below the values for each other country.
Note that for the vast majority of the countries skewness and kurtosis for the distribution of $\nu_i$ are in the vicinity of 0 and 3, respectively (which are the values one would expect for normal distributions).
However, the moments of the distributions do depend on the data aggregation level.
Fig.\ref{MomentAggregate} shows skewness and kurtosis for the distributions of $\nu_i$ for each election on each aggregation level.
By increasing the data resolution, skewness and kurtosis for Russia decrease and approach similar values as observed in the rest of the countries, see also SI Tab.S3.
These measures depend on the data resolution and thus can not be used as unambiguous signals for statistical anomalies.
As will be shown, the fraud parameters $f_i$ and $f_e$ do {\it not} significantly depend on the aggregation level or total sample size.

\subsection{Voting model results}

Estimation results for $f_i$ and $f_e$ are given in Tab.S3 for all countries on each aggregation level.
They are zero (or almost zero) in all of the cases except for Russia and Uganda.
In the right column of Fig.\ref{Figure2} we show the model results for Russia (2011 and 2012), Uganda and Switzerland for $f_i=f_e=0$.
The case where both fraud parameters are zero corresponds to the absence of incremental and extreme fraud mechanisms in the model and can be called the 'fair election' case.
In the middle column of Fig.\ref{Figure2} we show results for the estimated values of $f_i$ and $f_e$. 
The left column shows the actual vtd of the election.
Values of $f_i$ and $f_e$ significantly larger than zero indicate that the observed distributions may be affected by fraudulent actions.
To describe the smearing from the main peak to the upper right corner which is observed for Russia and Uganda,  an incremental fraud probability around $f_i=0.64(1)$ is needed 
for {\em United Russia} in 2011 and $f_i=0.39(1)$ in 2012. This  means fraud in about 64\% of the units in 2011 and 39\% in 2012.
In the second peak close to 100\% turnout there are roughly 3,000 units  with 100\% of votes for United Russia in the 2011 data,
representing an electorate of more than two million people. 
Best fits yield  $f_e=0.033(4)$ for 2011 and $f_e=0.021(3)$ for 2012, i.e. two to three percent of all electoral units experience extreme fraud.
A more detailed comparison of  the model performance for the Russian parliamentary elections of 2003, 2007, 2011 and 2012 
is found in Fig.S2.
Fraud parameters for the Uganda data in Fig.\ref{Figure2} are found to be $f_i=0.49(1)$ and $f_e=0.011(3)$.
A best fit for the election data from Switzerland gives $f_i=f_e=0$.

These results are drastically more robust to variations of the aggregation level of the data than the previously discussed distribution moments skewness and kurtosis, Fig.\ref{FraudAggregate} and Tab.S3.
Even if we aggregate the Russian data up to the coarsest level of federal subjects (approximately 85 units, depending on the election), $f_e$ estimates are still at least two standard deviations
above zero, $f_i$ estimates more than ten standard deviations. 
Similar observations hold for Uganda.
For no other country, on no other aggregation level, such deviations are observed.
The parametric model yields similar results for the same data on different levels of aggregation, as long as the values maximizing the empirical vote (turnout) distribution and the distribution width remains invariant.
In other words, as long as units with similar vote (turnout) characteristics are aggregated to larger units, the overall shapes of the empirical distribution functions are preserved and the model estimates do not change significantly.
Note that more detailed assumptions about possible mechanisms leading to large heterogeneity in the data 
(such as the {\it Qu\'eb\'ecois} in Canada or voter mobilization in the Helsinki region in Finland, see SI) may have an effect 
on the estimate of $f_i$. 
However, these can under no circumstances explain the mechanism of extreme fraud. 
Results for elections in Sweden, UK and USA, where voting results are only available on a much coarser resolution ($\bar n_{pop} > 20000$), are given in Tab.S4.

Another way to visualize the intensity of election irregularities is the 
cumulative number of votes as a function of the turnout, Fig.\ref{Figure3}.
For each turnout level the total number of votes from units with this or lower levels are shown.
Each curve corresponds to the respective election winner in a different country, with average electorate per unit of comparable order of magnitude.
Usually these cdfs level off and form a plateau from the party's maximal vote count on.
Again this is not the case for Russia and Uganda. 
Both show a boost phase of increased extreme fraud toward the right end of the  distribution (red circles).
Russia  never even  shows a tendency to form a plateau.
As long as the empirical vote distribution functions remain invariant under data aggregation as discussed above, the shape of these cdfs will be preserved too.
Note that Fig.\ref{Figure3}. demonstrates that these effects are decisive for winning the 50\% majority in Russia 2011.

\section{Discussion}

We demonstrate that it is not sufficient to discuss the approximate normality of turnout, vote or logarithmic vote rate distributions, to decide if election results may be corrupted or not.
We show that these methods can lead to ambiguous signals, since results depend strongly on the aggregation level of the election data.
We developed a model to estimate parameters quantifying to which extent the observed election results can be explained by ballot stuffing.
The resulting parameter values are shown to be insensitive to the choice of the aggregation level.
Note that the error margins for $f_e$ values start to increase by decreasing $n$ below 100, see Fig.\ref{FraudAggregate}D, whereas $f_i$ estimates stay robust even for very small $n$.

It is imperative to emphasize that the shape of the fingerprints  in Fig.\ref{Figure1} 
will deviate from pure 2-d Gaussian distributions also as a result of non-fraudulent mechanisms, but due to heterogeneity in the population.
The purpose of the parametric model is to quantify to which extent ballot stuffing and the mechanism of extreme fraud may have contributed to these deviations, 
or if their influence can be ruled out on the basis of the data.
For the elections in Russia and Uganda they can not be ruled out.
As shown in Fig.S2, assumptions of their wide-spread occurrences even allow to reproduce the observed vote distributions to a good degree.

In conclusion it can be said with almost certainty that an election does not represent the will of the people, if 
a substantial fraction ($f_e$) of units reports a 100\% turnout with almost all votes for a single  
party, and/or if any significant deviations from the sigmoid form in the cumulative distribution of votes versus turnout are observed. 
Another indicator of systematic fraudulent or irregular voting behavior is an incremental fraud parameter $f_i$ which is significantly greater than zero on each aggregation level. 

Should such signals be detected it is tempting to invoke G.B. Shaw who held that 
"[d]emocracy is a form of government that substitutes election by the incompetent many for appointment by the corrupt few."





\section{Acknowledgments}
We acknowledge helpful discussions and remarks by Erich Neuwirth and Vadim Nikulin.
We thank Christian Borghesi for providing access to his election datasets and the anonymous referees for extremely valuable suggestions.






\pagebreak

\section{Supplementary Information}

\subsection{The data}
Descriptive statistics and official sources of the election results are shown in Tab.S1.
The raw data will be made available for download at http://www.complex-systems.meduniwien.ac.at/.
They report election results of parliamentary (Austria, Canada, Czech Republic, Finland, Russia, Spain and Switzerland), European (Poland) or presidential (France, Romania, Russia, Uganda) elections on at least one aggregation level.
In the rare circumstances where electoral districts report more valid ballots than registered voters, we work with a turnout of 100\%.
Territorial units with an electorate less than hundred are omitted at each point of the analysis, to avoid extreme vote and turnout rates as spurious results due to small communities.
The countries to include in this work have been chosen on the basis of data availability. 
A country is included, if the voting results are available in electronic form on an aggregation level where a number of $\bar n_{pop} \leq 5000$ vote eligible persons comprises one territorial unit.
Required data is the number of vote eligible persons $N_i$, the number of valid votes $V_i$ and the number of votes for the winning party/candidate $W_i$ for each unit.

\subsection{Model}
\begin{figure}[bp]
 \begin{center}
 \includegraphics[width=87mm]{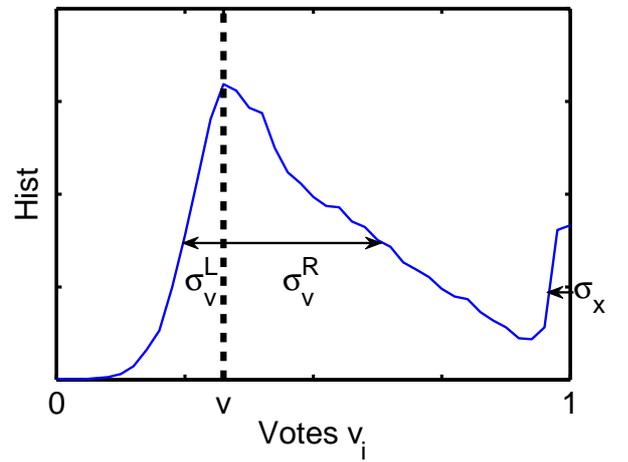}
  \end{center}
 \caption{A stylized version of an empirical vote distribution function shows how $v$, $\sigma_v^L$, $\sigma_v^R$ and $\sigma_x$ are estimated from the election results.
$v$ is the maximum of the distribution function. $\sigma_v^L$ measures the distribution width of values to the left of $v$, i.e. smaller than $v$.
The incremental fraud with $\sigma_v^R$ measures the distribution width of values to the right of $v$, i.e. larger than $v$.
The extreme fraud width $\sigma_x$ is the width of the peak at 100\% votes.}
  \label{SIFigureMeth}
 \end{figure}

A country is separated into $n$ electoral units $i$, each having an electorate of $N_i$ people and in total $V_i$ valid votes. 
The fraction of valid votes for the winning party in unit $i$ is denoted $v_i$.
The average turnout over all units, $\bar a$, is given by $\bar a = 1/n \sum_i (V_i/N_i)$ with standard deviation $s_a$,
the mean fraction of votes $\bar v$ for the winning party is $\bar v = 1/n \sum_i v_i$ with standard deviation $s_v$.
The mean values $\bar a$ and $\bar v$ are typically close to but not identical to the values which maximize the empirical distribution function of turnout and votes over all units.
Let $v$ be the number of votes where the empirical distribution function assumes its (first local) maximum (rounded to entire percents), see Fig. S\ref{SIFigureMeth}.
Similarly $a$ is the turnout where the empirical distribution function of turnouts $a_i$ takes its (first local) maximum.
The distributions for turnout and votes are extremely skewed to the right for Uganda and Russia which also inflates the standard deviations in these countries, see Tab. S2.
To account for this a 'left-sided' ('right-sided') mean deviation $\sigma_v^L$ ($\sigma_v^R$) from $v$ is introduced.
$\sigma_v^R$ can be regarded as the {\it incremental fraud width}, a measurable parameter quantifying how intense the vote stuffing is.
This contributes to the 'smearing out' of the main peaks in the election fingerprints, see Fig.1 in the main text.
The larger $\sigma_v^R$, the more inflated the vote results due to urn stuffing, in contrast to $\sigma_v^L$ which quantifies the scatter of the voters' actual preferences.
They can be estimated from the data by
\begin{eqnarray}
\sigma_v^L & = & \sqrt{ \langle (v_i-v)^2  \rangle_{v_i < v} } \quad, \\ 
\sigma_v^R & = &  \sqrt{ \langle (v_i-v)^2  \rangle_{v_i > v} } \quad.
\label{SigmaV}
\end{eqnarray}
Similarly the {\it extreme fraud width} $\sigma_x$ can be estimated, i.e. the width of the peak around 100\% votes.
We found that $\sigma_x=0.075$ describes all encountered vote distributions reasonably well.
For a visualization of  $\sigma_v^L$, $\sigma_v^R$ and $\sigma_x$ see Fig. S\ref{SIFigureMeth}.

While $f_i$ and $f_e$ measure in how many units incremental and extreme fraud occur, $\sigma_v^R$ and $\sigma_x$ quantify how intense these activities are, if they occur.
To get an estimate for the width of the distribution of turnouts over territorial unit which is free of possible fraudulent influences, the {\it turnout distribution width} $\sigma_a$ is calculated from electoral districts $i$ which have both $v_i<v$ and $a_i<a$, that is $\sigma_a=\sqrt{ \langle (a_i-a)^2  \rangle_{(a_i < a) \wedge (v_i < v)} }$.

Incremental fraud is a combination of two processes: stuffing ballots for one party into the urn and re-casting or deliberately wrong-counting ballots from other parties (e.g. erasing the cross).
Which one of these two processes dominates is quantified by the {\it deliberate wrong counting parameter} $\alpha>0$. 
For $0<\alpha<1$ the wrong-counting process dominates, for $\alpha>1$ the urn stuffing mechanism is prevalent.
In the following $\mathcal N(\mu, \sigma)$ denotes a normal distributed random variable with mean $\mu$ and standard deviation $\sigma$.
The model is specified by the following protocol, which is applied to each district.

\begin{itemize}
\item Pick a unit $i$ with electorate $N_i$ taken from the data.
\item The model turnout of unit $i$, $a^{(m)}_i$, is $\mathcal N(a, \sigma_a)$.
\item A fraction of $v_i^{(m)} \in \mathcal N(v,\sqrt(2) \sigma_v^L)$ people vote for the winning party.
\item With probability $f_i$ incremental fraud takes place. In this case the unit is assigned a fraud intensity $x_i \in | \mathcal N(0,\sqrt{\sigma_v^R})|$.
Values for $x_i$ are only accepted if they lie in the range $0<x_i<1$.
This is the fraction of votes not cast, $(1-a_i^{(m)})N_i$, which are added to the winning party.
Votes for the opposition are wrong counted for the winning party with a rate $x_i^\alpha$ (where $\alpha$ is an exponent).
To summarize, if incremental fraud takes place the winning party receives $N_i\left(a_i^{(m)} v_i^{(m)} + x_i (1-a_i^{(m)}) +x_i^{\alpha} (1-v_i^{(m)})a_i^{(m)}\right)$ votes.
\item With probability $f_e$ extreme fraud takes place. In this case opposition votes are canceled and added to the winning party with probability $y_i \in 1-|\mathcal N(1,\sigma_x)|$ (i.e. the above with $y_i$ replacing $x_i$).
Acceptable values for $y_i$ are again from the range $0<y_i<1$.
\end{itemize}

\subsubsection{Fitting the model}
The parameters for incremental and extreme fraud, $f_i$ and $f_e$, as well as the deliberate wrong counting parameter $\alpha$, are estimated by a goodness-of-fit test.
Let $pdf(v_i)$ be the empirical distribution function of votes for the winning party (the data is binned with one bin corresponding to one percent) over all territorial units.
The distribution function for the model units $pdf(v_i^{(m)})$ is calculated for each set of $(f_i,f_e,\alpha)$ values where $f_i,f_e \in \{0, 0.01, 0.02, \dots 1\}, \alpha \in \{0, 0.1 \dots 5\}$.
We report values for the fraud parameters where the statistic
\begin{equation}
 S(f_i,f_e,\alpha) = \sum_{i=1}^n \left( \frac{pdf(v_i)-pdf(v_i^{(m)})}{pdf(v_i)} \right) ^2 
\label{GOF}
\end{equation}
assumes its minimum, averaged over 100 realizations over the parameter space, see Tab.S3 for $f_i$ and $f_e$.

The extreme fraud parameter $f_e$ is zero (within one standard deviation) for almost all elections except Russia (2003, 2007, 2011 and 2012) and Uganda.
For very small $n$ ($n<100$) estimates for $f_e$ become less robust.
These are also the only elections where the incremental fraud parameter $f_i$ is not close to zero.
Values for $\alpha$ for the Russian elections are $\alpha_{Ru03} = 2.5(1)$ (2003), $\alpha_{Ru07} = 2.2(2)$ (2007), $\alpha_{Ru11} = 2.3(3)$ (2011), $\alpha_{Ru12} = 1.5(2)$ (2012), 
and $\alpha_{\mathrm{Uganda}} = 0.31(3)$ for Uganda.
Results for $\alpha$ from countries where $f_i$ is close to zero can not be detected in a robust way and are superfluous, since there are (almost) no deviations from the fair election case.

 Special care is needed in the interpretation of $f_i$ and $f_e$ values in countries where election units contain several polling stations.
 It may be the case that extreme fraud takes only in a subset of the polling stations within a unit place.
 In that case extreme fraud would be indistinguishable from the incremental fraud mechanism.

\begin{figure}[tb]
 \begin{center}
 \includegraphics[width=87mm]{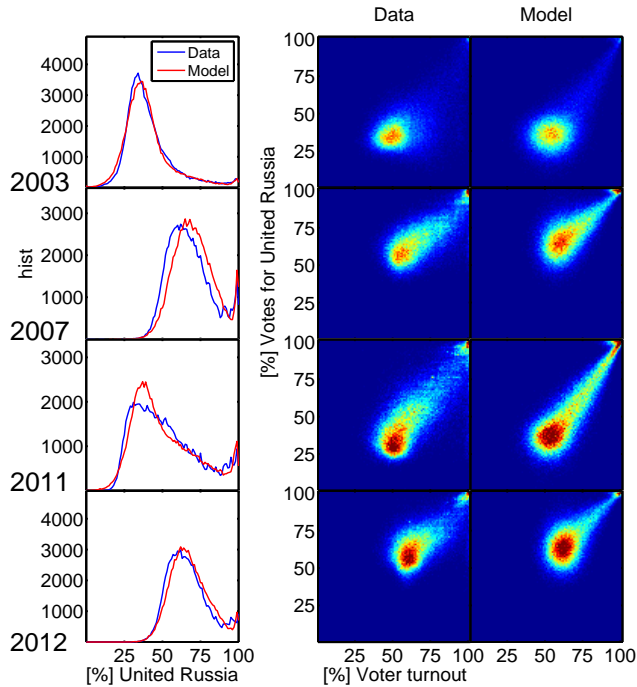}
  \end{center}
 \caption{Comparison of results from 2003, 2007, 2011 and 2012 Russian elections and the fraud model.
In the left column the distributions of the number of districts with a given percentage of votes for United Russia is shown for data (blue) and fraud model (red).
The middle column shows the observed vtds.
The data from 2007, 2011 and 2012 shows the same pattern, although the main cluster for United Russia is at a higher percentage of votes.
For 2003 there is a smaller number of districts with 100\% turnout and votes, and the main cluster is spread out to a smaller extent.
The right column shows fits for the data with the fraud model, using parameters $f_i = 0.31, f_e = 0.009$ (2003), $f_i=0.636, f_e = 0.038$ (2007),  $f_i=0.64, f_e=0.033$ (2011) and $f_i=0.39, f_e=0.021$ (2011).}
  \label{SIFigure2}
 \end{figure}

\subsection{On alternative  explanations for  election irregularities}
It is hard to construct other plausible mechanisms leading to a large number of territorial units having 100\%  turnout and votes for a single party than urn stuffing.
The case is not so clear for the 'smeared out' main cluster. 
In some cases, namely Canada and Finland, this cluster also takes on a slightly different form.
This effect clearly does not inflate the turnout as much as it is the case in Russia and Uganda, but it is nevertheless present.

In Canada the distribution of vote preferences is bimodal, with one peak around 50\% and one around 10\% (of the vote eligible population), but with similar turnout levels.
This is a result of a large-scale heterogeneity in the data: English and French Canada.
Votes are shown for the winning Conservatives. Looking at their results by province, they tallied 16.5\% votes cast in Quebec, but more than 40\% in eight of the remaining twelve other provinces.
As a consequence the logarithmic vote rate kurtosis becomes inflated.
However, these statistical deviations are perfectly distinguishable from the traces of ballot stuffing, resulting in vanishing fraud parameters on all aggregation levels.

Another possible mechanism leading to irregularities in the voting results is successful voter mobilization.
This may lead to a correlation between turnout and a party's votes.
The Finland elections, for example, where marked by radical campaigns by the True Finns.
They managed to mobilize evenly spread out across the country, with the exception of the Helsinki region, where the winning National Coalition Party performed significantly better than in the rest of the country.


\end{document}